\documentclass[aps,prl,twocolumn,groupedaddress,showpacs]{revtex4}

\usepackage{graphicx}
\usepackage{dcolumn}
\usepackage{bm}
\usepackage{color}

\begin{document}


\newcommand{\ket}[1]{|#1\rangle}

\title{Enhancing the area of a Raman atom interferometer using a versatile double-diffraction technique}

\author{T.~L\'{e}v\`{e}que}
\author{A.~Gauguet\footnote{Present address: Department of Physics, Durham University, Rochester Building, South Road, Durham DH1 3LE, England}}
\author{F.~Michaud}
\author{F.~Pereira Dos Santos}
\author{A.~Landragin}

\email[]{arnaud.landragin@obspm.fr}
\affiliation{LNE-SYRTE, UMR 8630 CNRS, UPMC, Observatoire de Paris, 61 avenue de l'Observatoire, 75014 Paris,
FRANCE}

\date{\today}

\begin{abstract}
In this paper we demonstrate a new scheme for Raman transitions which realize a symmetric momentum-space splitting of $4 \hbar k$,
deflecting the atomic wave-packets into the same internal state. Combining the advantages of Raman and Bragg diffraction, we
achieve a three pulse state labelled interferometer, intrinsically
insensitive to the main systematics and applicable to all kind of atomic sources. This splitting scheme can be
extended to $4N \hbar k$ momentum transfer by a multipulse
sequence and is implemented on a $8 \hbar k$ interferometer.
We demonstrate the area enhancement by measuring inertial forces.
\end{abstract}


\maketitle

Atom interferometers are of interest for precision measurements of
fundamental
constants~\cite{chu2002,clade2006,kasevich2007,lamporesi2008} and
for inertial measurements~\cite{peters2001,gustavson2000}. The
sensitivity of these apparatus is closely related to the ability
to diffract the atomic wave-packet in a coherent way. The angular
splitting of this process determines the enclosed area of the
interferometer and so its sensitivity. Many techniques have been
implemented to coherently separate atomic waves and increase the
angular deflection using material gratings~\cite{pritchard1991},
magneto-optical beam-splitters~\cite{pfau1993,schumm2005} or
momentum transfer by adiabatic passage~\cite{weitz1994}. For
precision measurements, the use of two photon transitions are the
most frequently used since they guarantee the control of the
momentum transfer during the diffraction process.

Matter-wave gratings made of a near resonant standing light
wave, permit the diffraction of the atoms along several coherent
paths, separated by $2 \hbar k$, in Kapitza-Dirac or Bragg regime. As in this kind of
interferometer~\cite{rasel1995,giltner1995,cahn1997} atoms always travel in
the same internal state, the output phase-shift is intrinsically
insensitive to many systematics such as the AC Stark shift and
temporal fluctuations of the Zeeman effect. These methods require
the use of highly collimated atomic sources for detection on the external degree of freedom or careful analysis for multiple path fringe pattern~\cite{cahn1997}. Alternatively, stimulated Raman
transitions~\cite{kasevich1991}, which split a wave packet into
two distinct internal states, are well adapted for use with
non-collimated atomic sources such as optical molasses. In
addition, they allow to realize interferometers which benefit from
an internal state labeling~\cite{borde1989} for the output phase
measurement, as populations of the output ports can be detected
using a state selective technique.

Many improvements have been made to increase the momentum-space
splitting of Bragg
~\cite{giltner1995,vigue2005,koolen2002,wang2005,gupta2002,muller2008}
and Raman~\cite{mcguirk2000} based interferometers. Recent works
have also permitted an enhancement in the momentum space splitting
by using conventional Raman and Bragg schemes coupled to Bloch
oscillations~\cite{clade2009,muller2009_2,dubetsky2002,salger2009}.

\begin{figure}[!h]
 \includegraphics[width=8.5cm]{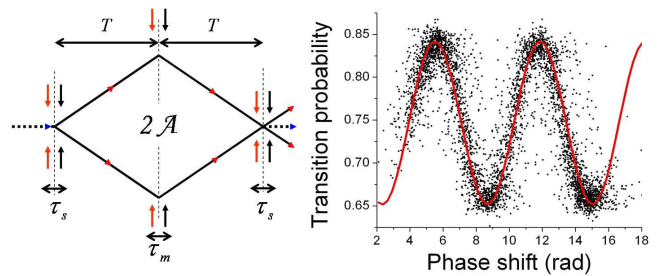}
 \caption{Left: Scheme of the $4 \hbar k$ interferometer realized
 using three double-diffraction pulses of duration $\tau_{s}$, $\tau_{m}$,
 $\tau_{s}$. Dotted and solid lines respectively represent wave packets
 in the state $\ket{g}$ and $\ket{e}$.
 Right: Interferometer signal for the source~$S_{1}$ as a
function of the phase-shift calculated from the tiltmeter signal.}
 \label{figure_3_pulse_franges}
 \end{figure}

Here, we demonstrate a new scheme based on Raman transitions,
which achieves a symmetric momentum-space splitting of $4 \hbar
k$~(Fig.~\ref{figure_3_pulse_franges}). It combines the advantages
of these two methods. First, starting with atoms from an optical
molasses, we realize an interferometer where partial wave packets
propagate along its two arms in the same internal state. Second,
the measurement of the output phase shift can still be realized
thanks to the state labeling. The area of the interferometer is
then increased by a factor of two compared to the traditional
$\pi/2$-$\pi$-$\pi/2$ three pulse configuration. This scheme can be easily extended to perform a~${4N\hbar k}$ momentum
splitting thanks to a multipulse sequence technique. As an example,
we utilize a $8 \hbar k$ interferometer for inertial measurements.

In our apparatus, described in~\cite{canuel2006}, $10^7$~Caesium
atoms are loaded from a vapor into two independent magneto-optical
traps. Two Caesium clouds are then launched at a temperature of
1.2~$\mu$K into two opposite parabolic trajectories using moving
molasses at velocity $v~=~2.4$~m.s$^{-1}$, oriented at an angle of
$8^\circ$ with respect to the vertical direction. At the apex of
their trajectory the atoms interact with horizontal Raman laser
pulses which act on the matter-wave as beam splitters or mirrors,
and generate an interferometer with $2T~=~60$~ms of total
interaction time. The use of two atomic
sources~($S_{1}$~and~$S_{2}$) launched along opposite directions
allows discrimination between the acceleration and
rotation~\cite{gustavson2000}.

Raman transitions are implemented in a retroreflected geometry.
The two laser beams of parallel polarizations at $\lambda=852$~nm
are guided to the setup through the same polarizing fiber. The
beams pass through the vacuum chamber and are reflected by a
mirror, crossing a quarter-wave plate twice. The wave plate is set
in such a way that the reflected lasers have orthogonal linear
polarization compared to the incident ones. As a result,
counterpropagating Raman transitions are allowed while
copropagating Raman transitions are forbidden. This commonly used
geometry limits the impact of wave-front aberrations and provides
a simple way to implement the $k$ reversal
technique~\cite{durfee2006}. In fact, in this scheme, atoms
interact with four laser waves which drive Raman transitions with
effective wave vectors ${\pm \mathbf{k_{eff}}=\pm
(\mathbf{k_{1}}-\mathbf{k_{2}})}$.

Atoms are initially prepared in the ground state
$\ket{6S_{1/2},F=3,m_{F}=0}$, which is written as $\ket{g}$. Our
four laser design, efficiently couples this initial state to the
excited state $\ket{6S_{1/2},F=4,m_{F}=0}$ by transferring a $2
\hbar k$ momentum along two opposite directions. The state
$\ket{g,p}$ is then coupled to the two states $\ket{e,p+2\hbar k}$
and $\ket{e,p-2\hbar k}$ where~$p$ is the initial momentum state
of the atom~\cite{canuel2006}. Due to the Doppler shift ${\omega_{D}=(p \cdot
k_{eff})/M}$, employing a retroreflected configuration only allows
the deflection of matter-waves in one momentum state or the other~\cite{peters2001,canuel2006,ice,pereira2007}.
In our new method, we take advantage of a null Doppler shift to
simultaneously couple the atoms in the two symmetric momentum
states. This double-diffraction scheme results in a momentum
splitting of $4 \hbar k$ with the same internal state~$\ket{e}$ in
the two paths.

\begin{figure}[!h]
 \includegraphics[width=5.5cm]{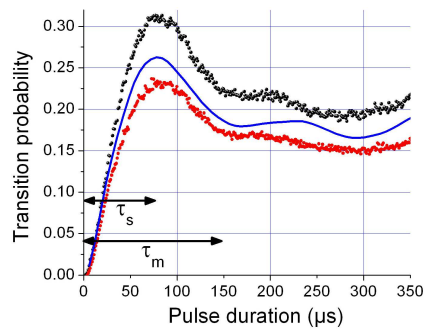}
 \caption{Rabi oscillation profiles obtained for the two atomic
  sources~$S_{1}$~and~$S_{2}$ at the apex of their trajectory (dots).
  The curve is the result of a numerical simulation.}
 \label{oscillations_rabi}
 \end{figure}

First, we investigate the efficiency of this beam splitter scheme
by comparing experimental results and numerical simulations of the
evolution of the atomic states, when interacting with the four
laser waves. The Fig.~\ref{oscillations_rabi} shows the
measurement of the transition probability from the ground state
$\ket{g}$ to the excited state $\ket{e}$ obtained for the atomic
clouds as a function of pulse duration. The effective Rabi
frequency, defined as usual~\cite{moler1992}, is
$\Omega_{eff}$~=~2.77~$\times$~10$^{4}$~rad.s$^{-1}$. Experimental
data are in good agreement with the results of a numerical
calculation, displayed as a thick line, in which the basis states
are restricted to five momentum states, adding the $\ket{g,{p\pm
4\hbar k}}$ detuned states to the three previous ones, $\ket{g}$ and $\ket{e,{p\pm
2\hbar k}}$. We verified that the population in the higher momentum states are negligible
taking into account our experimental parameters. We consider the
case in which the frequency difference between the lasers is set
to ${\delta\omega_{1}=(\omega_{e}-\omega_{g})+\omega_{R}}$ where
${\omega_{R}=(\hbar {k_{eff}}^{2})/2M}$ is the recoil frequency.
By extension of the calculation presented in~\cite{moler1992}, we
solve the Schr\"{o}dinger equation and calculate the evolution of
the populations. The calculation takes into account the Doppler
shift linked to the velocity dispersion of the atoms and the
spatial dependance of $\Omega_{eff}$ due to the expansion of the
atomic clouds in the gaussian profile of the Raman beam. In order
to keep only the atoms in the two symmetric states ${\ket{e,p \pm
2\hbar k}}$, a pushing beam is implemented after the Raman pulse
so as to remove remaining atoms in the ground state $\ket{g}$. In
addition, using this scheme, we have verified on the experiment
that the beam splitter also realizes a velocity selection of the
input momentum state (5.9~mm.s$^{-1}$ FWHM).

Using this double-diffraction scheme, we realize an interferometer
with a three pulse sequence~(Fig.~\ref{figure_3_pulse_franges}).
The first pulse of duration ${\tau_{s}=\pi/\sqrt{2}\Omega_{eff}}$
splits the input state $\ket{g,p}$ into ${\ket{e,{p\pm 2\hbar
k}}}$ symmetrically. The second pulse of duration
${\tau_{m}=\sqrt{2}\pi/\Omega_{eff}}$ acts as a mirror by coupling
each path to its opposite momentum state. Finally, matter-waves
are recombined thanks to a third pulse, of duration $\tau_{s}$, to
form a symmetric interferometer in which the matter-waves
propagate in the same internal state $\ket{e}$. A pushing beam is
implemented after each of the two first Raman pulses so as to
suppress spurious interferometers. An important feature of this Raman transition scheme 
is the use of state labeling at the output of the interferometer even if the matter-waves travel in the same
internal state. The output phase-shift can thus easily be measured
thanks to a fluorescence technique. The output atomic phase-shift
of our device is sensitive to acceleration $a$ and rotation
$\Omega$ as ${\Delta\Phi = 2\ \mathbf{k_{eff}}\cdot (\mathbf{a}  -
2 ( \mathbf{v} \times \mathbf{\Omega} )) T^2}$~\cite{canuel2006},
increasing the sensitivity of the traditional
$\pi/2$-$\pi$-$\pi/2$ three pulse geometry by a factor of two.

Similar to a Bragg interferometer, our apparatus is, in principle, insensitive to the main systematic effects: AC Stark shift, temporal fluctuations of the Zeeman effect  and magnetic field gradient along the direction of propagation of the atomic cloud. Moreover, the interferometer phase does not depend on the laser phase difference as its influence is identical over the two paths.

In order to scan the phase, a controlled acceleration phase shift
is induced on the interferometer. This is simply achieved by
tilting the apparatus ($\pm 50 \mu$rad) to get a small projection
of the gravity along the axis of the light-pulses. The inclination
is measured thanks to a calibrated tiltmeter~(Applied Geomechanics
701-2A). In Fig.~\ref{figure_3_pulse_franges} is displayed the transition
probability as a function of the phase shift deduced from the tilt
measurement. We observed a contrast of 20\% identical for the two
atomic sources, attributed to inhomogeneity of the Raman intensity over the atomic cloud.

\begin{figure}[!h]
 \includegraphics[width=8.5cm]{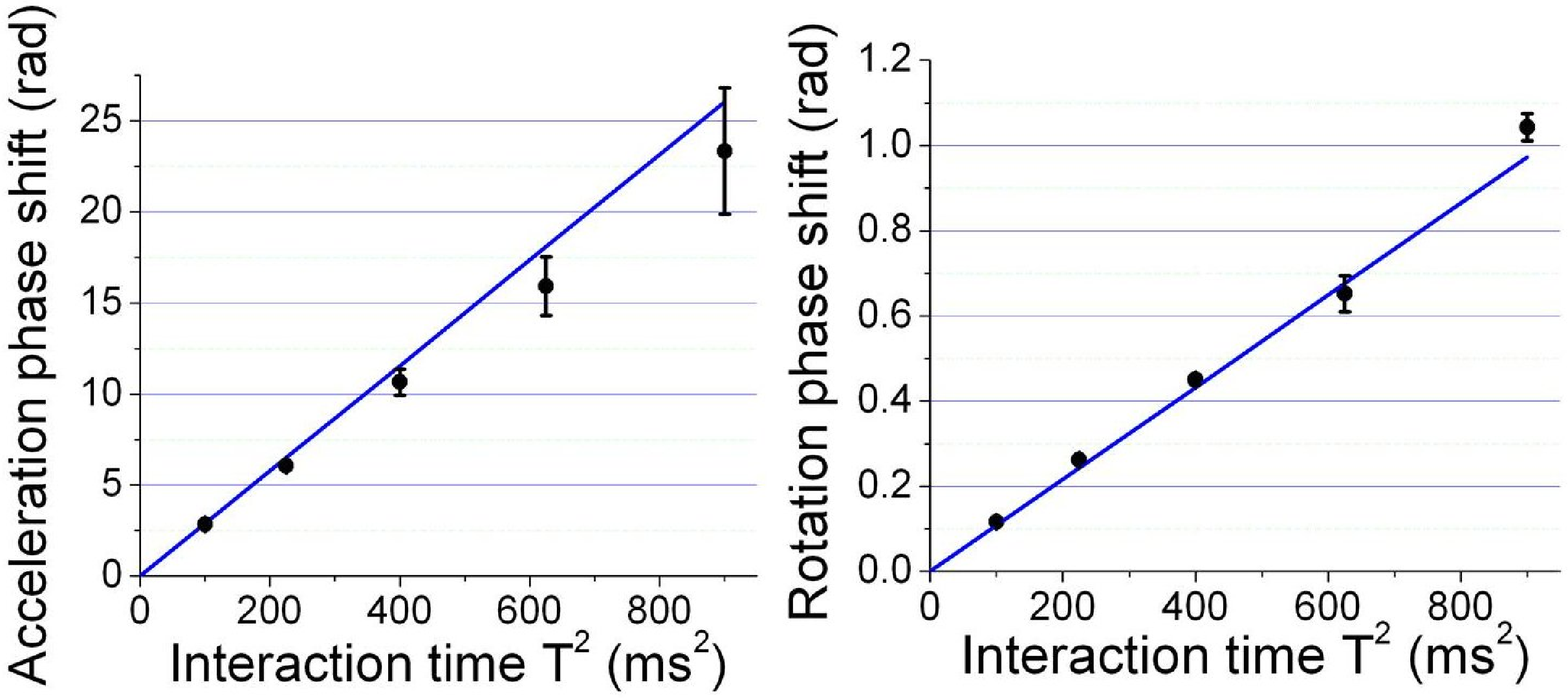}
 \caption{Quadratic scaling of the acceleration
phase shift induced by the projection of $g$ for an angle of
100~$\mu$rad with respect to the horizontal plane (Left) and of
the rotation phase-shift induces by the vertical component of the
rotation rate $\Omega^{E}_{z}$ (Right) as a function of the
interaction time $T^{2}$. The measurements (dots) are compared
with the expected behaviors (lines).}
 \label{mesures_3_pulses}
 \end{figure}

Using horizontal Raman beams makes the interferometer sensitive to
the vertical component of the Earth rotation rate
$\Omega^{E}_{z}$=$5.49 \times 10^{-5}$ rad.s$^{-1}$ at the
latitude of Observatoire de Paris. The quadratic scaling of the
rotation and acceleration phase shifts with the interaction time
$T$ is verified from $T$~=~10 to
30~ms~(Fig.~\ref{mesures_3_pulses}). The fits of experimental
phase shifts give measurements of the scale factors compatible
with the expected values, respectively to 3\% and 4\% for
acceleration and rotation. We deduce a rotation rate of
$\Omega_{z}$=$5.75 \times 10^{-5}$ rad.s$^{-1}$ in agreement with
the expected projection of the Earth's rotation rate within the
error bar.

The double-diffraction scheme can be extended to higher momentum
space splitting thanks to the retroreflected configuration. After
splitting the atoms into the two symmetric states $\ket{e,{p\pm
2\hbar k}}$ with the first pulse, a second pulse can be
implemented to efficiently couple these two resulting states to
the next momentum states $\ket{g,{p\pm 4\hbar k}}$. The Raman
frequency difference of the second pulse has to be changed to
${\delta \omega_{2}=(\omega_{e}-\omega_{g})-3 \omega_{R}}$ in
order to fulfill the resonance condition between the momentum
states and the two pairs of Raman lasers. This scheme can be
extended to a N pulse sequence to reach up to $4N \hbar k$
momentum space splitting. In order to conserve the resonance
condition, Raman frequency difference of the i$^{th}$ pulse has to
be changed and set to ${\delta \omega_{i}=(\omega_{e}-\omega_{g})+
(-1)^{i-1} (2i-1)\omega_{R}}$. Following this argument, symmetric
Raman transitions are suitable to realize a large area
interferometer with $4 N \hbar k$ momentum space splitting.

\begin{figure}[!h]
 \includegraphics[width=5.5cm]{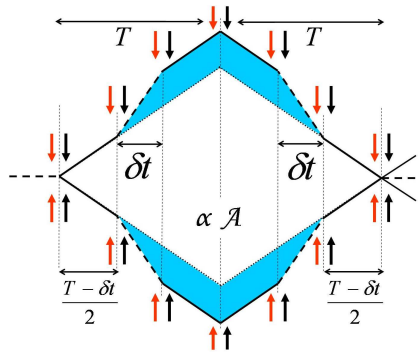}
 \caption{Recoil diagram for the $8 \hbar k$
 interferometer reaching a total area enhanced by a factor of~$\alpha$.
 Dotted lines represent wave packets in the $\ket{g}$
 hyperfine ground state while solid lines depict the $\ket{e}$ ground
 state. Colored surface represents the area enhancement compared
 with the corresponding three pulses configuration.
 }
 \label{interferometre_7_pulses}
 \end{figure}

Here, we demonstrate an interferometer realized with $N=2$,
reaching a $8 \hbar k$ momentum space splitting between the two
paths, using seven Raman pulses. In this
configuration~(Fig.~\ref{interferometre_7_pulses}), the first
atomic beam splitter is composed of two light-pulses respectively
set with detunings $\delta \omega_{1}$ and $\delta \omega_{2}$. In
this manner, matter-waves are coupled from the state $\ket{g,p}$
to $\ket{g,{p\pm 4\hbar k}}$. The mirror is realized by three
successive pulses at $\delta \omega_{2}$,~$\delta
\omega_{1}$,~$\delta \omega_{2}$ in order to deflect the atoms in
their opposite momentum-state. Finally, the two interferometric
paths are recombined with the symmetric $\delta
\omega_{2}$,~$\delta \omega_{1}$ light-pulse sequence. The
durations of the seven pulses are respectively
$\tau_{s}$,~$\tau_{\pi}$,~$\tau_{\pi}$,~$\tau_{m}$,~$\tau_{\pi}$,~$\tau_{\pi}$,~$\tau_{s}$,
where $\tau_{\pi}=\pi/\Omega_{eff}$. In addition, four pushing
pulses are applied to blow away residual atoms remaining in the
internal state $\ket{g}$.

\begin{figure}[!h]
 \includegraphics[width=8.5cm]{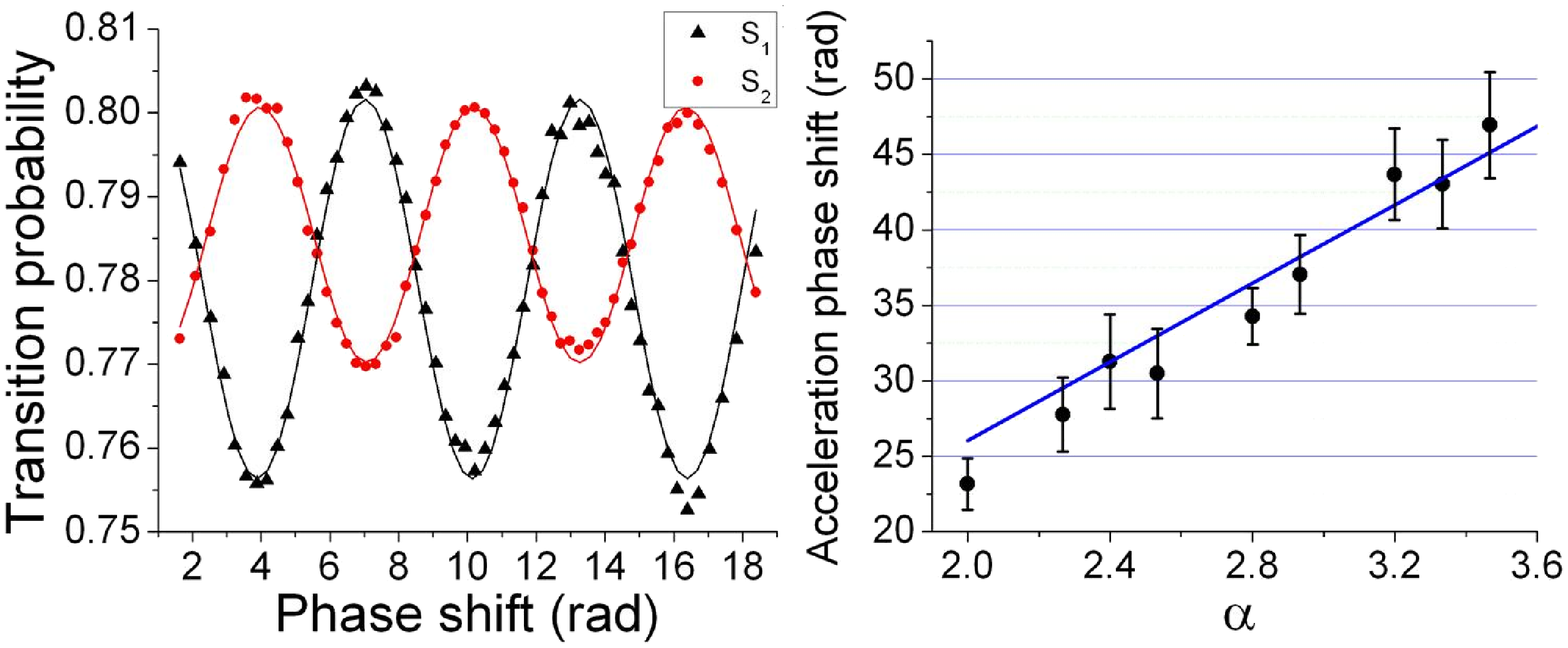}
 \caption{Left: Fringe patterns averaged over 450 samples.
 Right: Linear scaling of the acceleration phase-shift induced by
the projection of $g$ for an angle of 100~$\mu$rad with respect to
the horizontal plane
 as a function of the area enhancement factor $\alpha$. The measurement (dots) is compared with the
expected values (line).}
 \label{mesures_7_pulses}
 \end{figure}

The total interaction time of this interferometer is ${2T=60}$~ms.
Atoms propagates in the states $\ket{g,{p\pm 4\hbar k}}$ during
$2\delta t=40$~ms with a delay of ${(T-\delta t)/2}$. The
interferometric area increases by a factor of ${\alpha = 2(1 +
\delta t/{T})}$ compared with the $\pi/2-\pi-\pi/2$ configuration.
The sensitivity to inertial effects is then proportionally
enhanced up to ${\Delta\Phi = \alpha\ \mathbf{k_{eff}}\cdot
(\mathbf{a}  - 2 ( \mathbf{v} \times \mathbf{\Omega} )) T^2}$. As
before, this phase is scanned by tilting the interferometer. In
Fig.~\ref{mesures_7_pulses} we show the fringe patterns of the two
sources, obtained with the $8 \hbar k$ interferometer. The phase
shift between the two interferometers corresponds to the rotation
phase-shift induced by the vertical component of the Earth
rotation rate $\Omega^{E}_{z}$. The linear scaling of the
acceleration phase-shift with $\alpha$ was investigated by
changing $\delta t$. Figure~\ref{mesures_7_pulses} displays the
result of the measurement for $\delta t=0$~to~20~ms which shows a
good agreement with the expected slope within 2\%. This
corresponds to an enhanced sensitivity of a factor of 3.46
compared to the usual $\pi/2-\pi-\pi/2$ configuration.

With respect to the $4 \hbar k$ case of a three pulse
interferometer, the atoms do not remain into the same internal
state all along the two paths. Nevertheless, as the internal state
changes at the same time for the two partial wave packets, the
enhanced interferometer remains insensitive to the same
systematics as in the Bragg case.

To conclude, we demonstrated a new method of achieving symmetric
Raman transitions with $4 \hbar k$ momentum-space splitting,
deflecting the atoms into the same internal state. Using this
double-diffraction technique, we have realized an interferometer
of increased area which has been tested by performing acceleration
and rotation measurements. The splitting scheme has been extended
to reach a ${8 \hbar k}$ interferometric geometry in a seven pulse
configuration, opening the way for enhanced ${4N \hbar k}$ path
separation geometry.

This scheme provides a simple way to increase the sensitivity of
the apparatus without any significant changes in the experimental
setup thanks to the retroreflected Raman configuration. In
particular, it does not require extra laser power compared to
standard Raman transition based interferometers, in contrast with other methods. Furthermore, as
matter-waves are symmetrically driven in the same ground state,
interferometers are intrinsically insensitive to the AC Stark
shift, the Zeeman effect and Raman laser phase-noise. The enhanced
area geometry thus combines the advantages of Raman and Bragg
based interferometers. Moreover, the principle of the
double-diffraction can be extended to the case of a non-zero
Doppler shift, providing that the two opposite Raman transitions
remain resonant simultaneously. This general case require at least
three different laser frequencies.

Finally, this work provides insight into space-atom interferometer
design~\cite{ice}. Indeed, for inertial sensors in zero-gravity
environments, Doppler effect can not be used to select one or the
other effective Raman transitions in the current retroreflected
configuration needed for accuracy~\cite{pereira2007}. The
symmetric Raman diffraction offers a promising solution to
circumvent this constraint.

\begin{acknowledgments}
We would like to thank the Institut Francilien pour la Recherche
sur les Atomes Froids (IFRAF) and the European Union (FINAQS
STREP/NEST project contract no 012986 and EuroQUASAR/IQS project)
for financial support. T.L. thanks the DGA for supporting his
work. We also thank P. Bouyer for careful reading.
\end{acknowledgments}


\end{document}